\documentclass[journal]{IEEEtran}

\usepackage{booktabs}
\usepackage{xcolor}
\usepackage{colortbl}
\usepackage{amssymb}
\usepackage{algorithmic}
\usepackage{amsmath} 
\usepackage{amsfonts}

\ifCLASSINFOpdf
\usepackage[pdftex]{graphicx}
\else
\usepackage[dvips]{graphicx}
\fi
\usepackage{epstopdf}
\usepackage{array}
\ifCLASSOPTIONcompsoc
\usepackage[caption=false,font=normalsize,labelfont=sf,textfont=sf]{subfig}
\else
\usepackage[caption=false,font=footnotesize]{subfig}
\fi
\usepackage{float} 
\usepackage{fixltx2e}
\usepackage{color}
\usepackage{xcolor}
\usepackage{textcomp}
\usepackage[normalem]{ulem}
\usepackage{nomencl}
\makenomenclature
\usepackage{etoolbox}
\renewcommand\nomgroup[1]{%
	\item[\bfseries
	\ifstrequal{#1}{P}{Physical Variables}{%
		\ifstrequal{#1}{M}{Mathematical Symbols}{%
			\ifstrequal{#1}{S}{Subscripts and Superscripts}{}}} ]}

\usepackage{hyperref}
\hypersetup{
	colorlinks=true,
	linkcolor=blue,
	filecolor=magenta,      
	urlcolor=cyan,}
\IEEEoverridecommandlockouts

\usepackage{amsmath}
\usepackage{amssymb}
\usepackage{graphicx}

\usepackage[
    backend=biber,
    style=ieee,
    sorting=none,
    doi=false,
    url=false,
    eprint=false,
    isbn=false
]{biblatex}
\DeclareFieldFormat{shortjournal}{\mkbibemph{#1}}
\renewbibmacro*{journal}{%
  \iffieldundef{shortjournal}
    {\printfield{journaltitle}}
    {\printfield{shortjournal}}
}
\begin{document}

\title{Stability-Constrained Imitation Learning for Performance-Guaranteed Neural Feedback Frequency Control in Microgrids}

\author{Sajid Kamal,~\IEEEmembership{Student Member, IEEE,}
Yichen Zhang,~\IEEEmembership{Senior Member, IEEE,}
Feng Qiu,~\IEEEmembership{Fellow, IEEE,}
Tianqiao Zhao,~\IEEEmembership{Member, IEEE}
\thanks{
S. Kamal, Y Zhang, and T. Zhao are with University of Texas at Arlington (e-mail: sajid.kamal@uta.edu, yichen.zhang@uta.edu, tianqiao.zhao@uta.edu). F. Qiu is with Argonne National Laboratory (e-mail: fqiu@anl.gov).}
}

\maketitle
\begin{abstract}
Guaranteeing performance for frequency control is challenging within traditional control paradigms, as it demands precise disturbance estimation and real-time control signal computation within a timeframe of seconds. This paper addresses this challenge by proposing a stability-certified imitation learning framework for frequency regulation in low-inertia microgrids. Based on the frequency response model of the microgrid, an offline numerical optimal control is employed to generate near-optimal pairs of state and control dataset, which are subsequently used to train a deep neural network (DNN) feedback controller. To ensure the closed-loop stability, the DNN controller is trained with a set of Lyapunov-based stability constraints derived using absolute stability theory. Such an imitation learning paradigm can significantly improve the training efficiency, while the resulting DNN controller acts as an end-to-end function that simultaneously handles disturbance estimation and control optimization. The proposed framework is validated on a modified IEEE 33-node diesel–wind microgrid. The DNN controllers are deployed in wind turbine generators (WTGs) to provide decentralized, “intelligent" supplementary active power support to always maintain frequency deviations within predefined safety bounds under different disturbances while preserving closed-loop stability.
\end{abstract}

\begin{IEEEkeywords}
Imitation learning, physics-constrained learning, frequency regulation, wind turbine generator (WTG), Lyapunov stability, linear matrix inequalities (LMIs).
\end{IEEEkeywords}

\section{Introduction}

The rapid integration of inverter-based resources (IBRs) has significantly degraded the frequency response of modern power systems \cite{guPowerSystemStability2023,zhangHybridControllerWind2017}, especially in IBR-dominated microgrids.
Since original IBRs are not designed to support grid frequency, a disturbance can lead to faster and deeper frequency deviations when synchronous generators are replaced by IBRs. 
To maintain frequency stability in such low-inertia environments, IBRs are expected to provide fast frequency support during inertial and primary response timeframes.

Among different IBRs, wind turbine generators (WTGs) become particularly attractive for frequency support due to the kinetic energy stored in their rotating masses. WTGs can temporarily extract this kinetic energy and inject fast active power into the grid by modulating the rotor-side converter (RSC) without additional deployment of energy storage \cite{liOptimalShortTermFrequency2026}. This capability makes WTGs particularly attractive for frequency support in low-inertia microgrids. However, designing frequency control for WTGs that can keep grid frequency deviations within predefined limits remains a challenging task~\cite{uriarteMicrogridRampRates2015,zhangHybridControllerWind2017,jiangFrequencyConstrainedDispatch2025,briseboisWindFarmInertia2011}. The underlying challenge of \emph{such a performance-guaranteed frequency control} is the requirement of precise disturbance estimation and real-time control signal computation within a timeframe of seconds. 

Frequency supportive control of IBRs can be categorized into pre-defined control, reactive control, and learning-based control. Pre-defined control employs a time-invariant control structure, such as droop \cite{vidyanandanPrimaryFrequencyRegulation2013,yangAdaptiveFrequencyDroop2024,zhangEffectsDroopBased2024}, traditional proportional-integral-derivative (PID) \cite{morrenWindTurbinesEmulating2006}, intelligent PID \cite{parkModelfreeControlFrequency2021}, power surge logic \cite{keungKineticEnergyWindTurbine2009,wangNovelActivePower2018}, fuzzy logic \cite{bevraniIntelligentFrequencyControl2012,zhangFuzzyLogicBasedFrequency2016,karimiInertiaResponseImprovement2020}, state and output feedback \cite{ghoshDoublyFedInduction2016,liuESOBasedInertiaEmulation2017,zhangProvisionGuaranteedInertial2018,morovatiRobustOutputFeedback2021}. The classic frequency support function with a WTG is presented in \cite{morrenWindTurbinesEmulating2006}, where a droop control admits the primary support and a derivative control realizes the inertia emulation. Classic PID designs compensation terms utilizing explicit model information, whereas an intelligent PID is proposed in \cite{parkModelfreeControlFrequency2021} to eliminate this requirement.
A variable droop control for primary frequency support is proposed in \cite{vidyanandanPrimaryFrequencyRegulation2013} to count for time-varying availability of wind energy. 
The frequency deviation-based droop control with dynamic gains is proposed in \cite{yangAdaptiveFrequencyDroop2024} to address the conflicting impact from the maximum power point tracking on frequency support. It has been found that a static derivative control can exhibit time-varying synthetic inertia constant \cite{wangQuantifyingSyntheticInertia2018}. Then, a model reference control is proposed to address this issue \cite{zhangProvisionGuaranteedInertial2018}, which is essentially a dynamic state feedback control using H-$\infty$ control. The linear-quadratic regulator (LQR) method can also be used to design state feedback \cite{markovicLQRBasedAdaptiveVirtual2019}. Fuzzy logic is proposed to enhance the frequency support performance by online tuning the PID control parameter in \cite{bevraniIntelligentFrequencyControl2012} and adding supplementary signal to reduce transients in \cite{karimiInertiaResponseImprovement2020}. As an open-loop logic, power surge control is proposed \cite{keungKineticEnergyWindTurbine2009,wangNovelActivePower2018} to avoid potential instability from the frequency support function. The advantages of pre-defined control is the negligible computation time for control signal, while the downside is the lack of capability to incorporate constraints, and therefore, cannot guarantee performance against different disturbances. 

Under the reactive control paradigm, control actions are computed in real time via optimal control frameworks, most notably through model predictive control (MPC) \cite{kouNonlinearModelPredictive2019,stanojevMPCBasedFastFrequency2022,liuCoordinatedDistributedMPC2017,liuTubeBasedDistributedMPC2024}. Due to the computation requirement, MPC is generally employed in secondary frequency control timeframe \cite{venkatDistributedMPCStrategies2008,liuCoordinatedDistributedMPC2017,liuTubeBasedDistributedMPC2024}. To address this challenge, finite-time MPC is employed in \cite{kouNonlinearModelPredictive2019}, whereas explicit MPC is employed in \cite{stanojevMPCBasedFastFrequency2022} for computing optimal control actions without the need for executing an optimization algorithm in real time. Both methods enable the MPC's application in primary frequency support. An approximated feedback is constructed based on optimal frequency trajectory in \cite{zhangSystemViewOptimalAdditional2024}. Another method is to use a near-real-time scheduling and real-time execution paradigm proposed in \cite{zhangRobustTrajectoryconstrainedFrequency2023}, where the nonlinear system is converted into a linear system with uncertainty bounds to count for linearization error. The advantage of reactive control is their ability to incorporate constraints. However, achieving strict performance guarantees remains contingent upon accurate modeling, precise disturbance estimation, and real-time execution of the control optimization framework.

Strictly speaking, learning-based control can be considered a subset of predefined control methods, given that most frameworks employ an offline-training, online-execution workflow. Nonetheless, recent breakthroughs in deep neural architectures and real-time online learning merit its treatment as an independent category. Learning-based frequency control has recently attracted significant attention due to its ability to provide near-instantaneous online inference for control signal and nonlinear control policies for strong adaptivity \cite{chenReinforcementLearningSelective2022}. Most works employ the reinforcement learning framework. The key challenge is to ensure the closed-loop stability under the neural feedback. One way is to filter out control actions that can lead to unstable system responses. A gauge map is constructed using set theory to provide safety guarantees \cite{tabasComputationallyEfficientSafe2022}. The distributional reinforcement learning is proposed in \cite{xieDistributionalDeepReinforcement2022} to achieve a more stable learning process. The control barrier function is embedded in the reinforcement learning in \cite{wanAdapSafeAdaptiveSafeCertified2023,zhaoBarrierCertificatedReinforcementLearning2023} to certify stable control actions. Another method is to ensure closed-loop stability by construction. Ref. \cite{cuiReinforcementLearningOptimal2023} shows the equilibrium's local stability is guaranteed if the control is monotone, and then constructs a monotone neural network. Ref. \cite{yuanReinforcementLearningDistributed2024} first certifies the region of control outputs that ensures stability and then designs an output-constrained neural network. Other works focus on multi-agent paradigm \cite{yanMultiAgentDeepReinforcement2020,chenModelFreeEmergencyFrequency2021,yangDistributedDynamicInertiaDroop2023}, secondary frequency control \cite{yanMultiAgentDeepReinforcement2020,adibiSecondaryFrequencyControl2022}, attention mechanism \cite{yangDataDrivenLoadFrequency2023}. However, reinforcement learning frameworks suffer from the well-known exploration and exploitation dilemma \cite{kangPolicyOptimizationDemonstrations2018}. A well-behaved controller may require significant numbers of samples, which can introduce high computation complexity considering the interaction with power system environments \cite{zhangHybridImitationLearning2022}.

To address the shortcomings of predefined and reactive control while improving the training efficiency of learning-based approaches, this paper proposes an imitation learning-based control paradigm for WTG frequency support in microgrids. Different from reinforcement learning that explores the action space stochastically, imitation learning employs expert knowledge to provide near-optimal control actions \cite{leHierarchicalImitationReinforcement2018,kangPolicyOptimizationDemonstrations2018}. 
Utilizing the augmented system frequency response (ASFR) model, where the impact from both disturbance and WTG's support on system frequency is modeled \cite{zhangRobustTrajectoryconstrainedFrequency2023}, a numerical optimal control problem is formulated as the expert so that near-optimal state and control pairs can be obtained to train a neural feedback controller. To ensure closed-loop stability, the paper proposes a \emph{physics-constrained training paradigm}, where the stability conditions are formulated as linear matrix inequalities (LMIs) based on the absolute stability theory \cite{yinImitationLearningStability2022}. The main contributions of this paper are as follows:
\begin{enumerate}
\item An imitation learning-based control paradigm is proposed for WTG frequency support in microgrids. Imitation learning establishes a powerful synergy between predefined and reactive control paradigms by embedding offline reactive control capabilities and online disturbance estimation into a predefined structure.
\item A stability-constrained neural feedback control training program is formulated using sector-bounded deep neural network (DNN) representations and Lyapunov-based LMI constraints to ensure closed-loop stability of the ASFR model.
\item A decentralized intelligent feature of the proposed method is validated using the electromagnetic (EMT) simulation on a modified IEEE 33-node diesel–wind microgrid under different disturbance conditions, demonstrating that frequency deviations can be maintained close to predefined safety limits even under a multi-machine scenario.
\end{enumerate}

The rest of the paper is organized as follows. Section II describes the control objective and workflow. Section III presents the microgrid ASFR model and numerical optimal control problem  as the expert. Section IV details the Lyapunov stability analysis and convex LMI formulation for DNN training. Section V discusses the control performance under the ASFR model. Section VI presents the case study validation on the 33-node microgrid, and conclusions are provided in Section VII.

\section{Problem Statement}
\subsection{Control Objective}
Let the microgrid ASFR be represented using a discrete-time state-space model:
\begin{equation}
x(k+1)=Ax(k)+Bu(k)+Ed(k), \qquad d(k)\in\mathcal{D}
\label{eq:system}
\end{equation}
where $x(k)$ represents the frequency-related system states, $u(k)=\pi(x(k);\theta)$ denotes the DNN-based supplementary frequency support signal generated by WTGs, and $d(k)$ represents disturbances such as sudden load changes, assumed to belong to the bounded disturbance set $\mathcal{D}$. The detailed model will be introduced in Section  \ref{subsec_ASFR}.

The goal is to develop a real-time DNN feedback control $u(k)=\pi(x(k);\theta)$ which provides supplementary frequency support for the WTGs under bounded but unknown disturbances while satisfying the following two requirements.

\subsubsection{Frequency Safety and Operational Constraints}
The controller must maintain the system frequency within prescribed safety limits during disturbance:
\begin{equation}
\overline{\Delta f}\geq \Delta f(k)\geq \underline{\Delta f},
\quad
\forall k\geq0,\; d(k)\in\mathcal{D}
\end{equation}
where $\overline{\Delta f}$ and $\underline{\Delta f}$ denote the maximum and minimum allowable frequency threshold and $\mathcal{D}$ represents the bounded disturbance set. In addition, all system states and control inputs must remain within their admissible operating limits:
\begin{equation}
x(k)\in\mathcal{X}_s,\qquad
u(k)\in\mathcal{U},
\qquad
\forall k\geq0
\end{equation}

\subsubsection{Closed-Loop Stability Requirement}
The DNN policy $u(k)=\pi(x(k);\theta)$ must guarantee the stability of the closed-loop system \ref{eq:system}. The training algorithm will search for a quadratic Lyapunov function  $V(x)=x^{\top}Px$ such that:
\begin{equation}
V(x(k+1))-V(x(k))<0
\end{equation}

\subsection{Workflow}
\begin{figure}[t]
    \centering
    \includegraphics[width=\columnwidth]{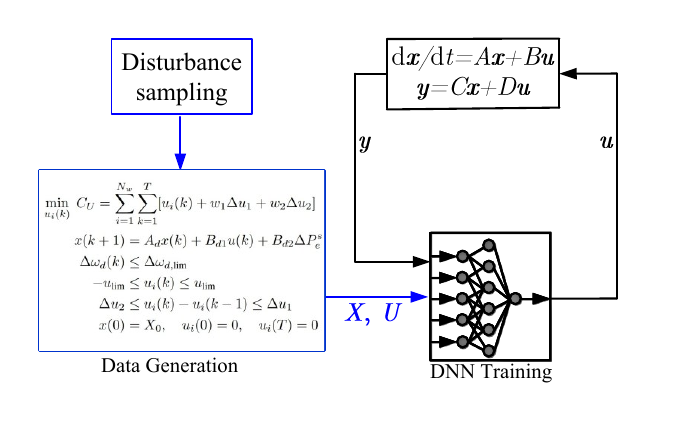}
    \caption{Overall workflow of the proposed imitation-learning-based frequency controller.}
    \label{fig_workflow}
\end{figure}

This section presents the overall workflow used to develop and validate the proposed DNN-based frequency controller for a diesel–wind microgrid. The workflow integrates microgrid modeling, offline data generation, and learning-based control within a comprehensive framework, as shown in Fig. \ref{fig_workflow}. The individual stages of the workflow are discussed below.

\subsubsection{Microgrid Frequency Response Model} First, the ASFR model will be derived from the full-scale nonlinear diesel–wind microgrid model to represent frequency dynamics under both disturbances and control actions. The microgrid includes diesel generators modeled as combustion engine-driven synchronous machines and type-3 wind turbine generators. This is discussed in detail in Section \ref{subsec_ASFR}.

\subsubsection{Offline Data Generation Using Numerical Optimal Control} Using the microgrid ASFR model, offline data are generated under various disturbance conditions. For each sampled disturbance, an offline numerical optimal control computes effective control actions that regulate frequency within the permissible ranges. Then, the system frequency and the corresponding control actions are recorded. This dataset captures necessary control from WTGs to maintain adequate frequency under different disturbances and serves as the training dataset.

\subsubsection{DNN Training and Closed-Loop Verification} Finally, a closed-loop stability condition is derived as LMI and embedded into the deep neural network training using imitation learning \cite{yinImitationLearningStability2022, fazlyabSafetyVerificationRobustness2022}. After training, the closed-loop stability and performance guarantee capability of the DNN controller is verified using the full-scale nonlinear microgrid simulation.

\section{Modeling and Control}
\subsection{Augmented System Frequency Response Model} \label{subsec_ASFR} 
\begin{figure*}[t]
	\centering
	\includegraphics[width=\textwidth]{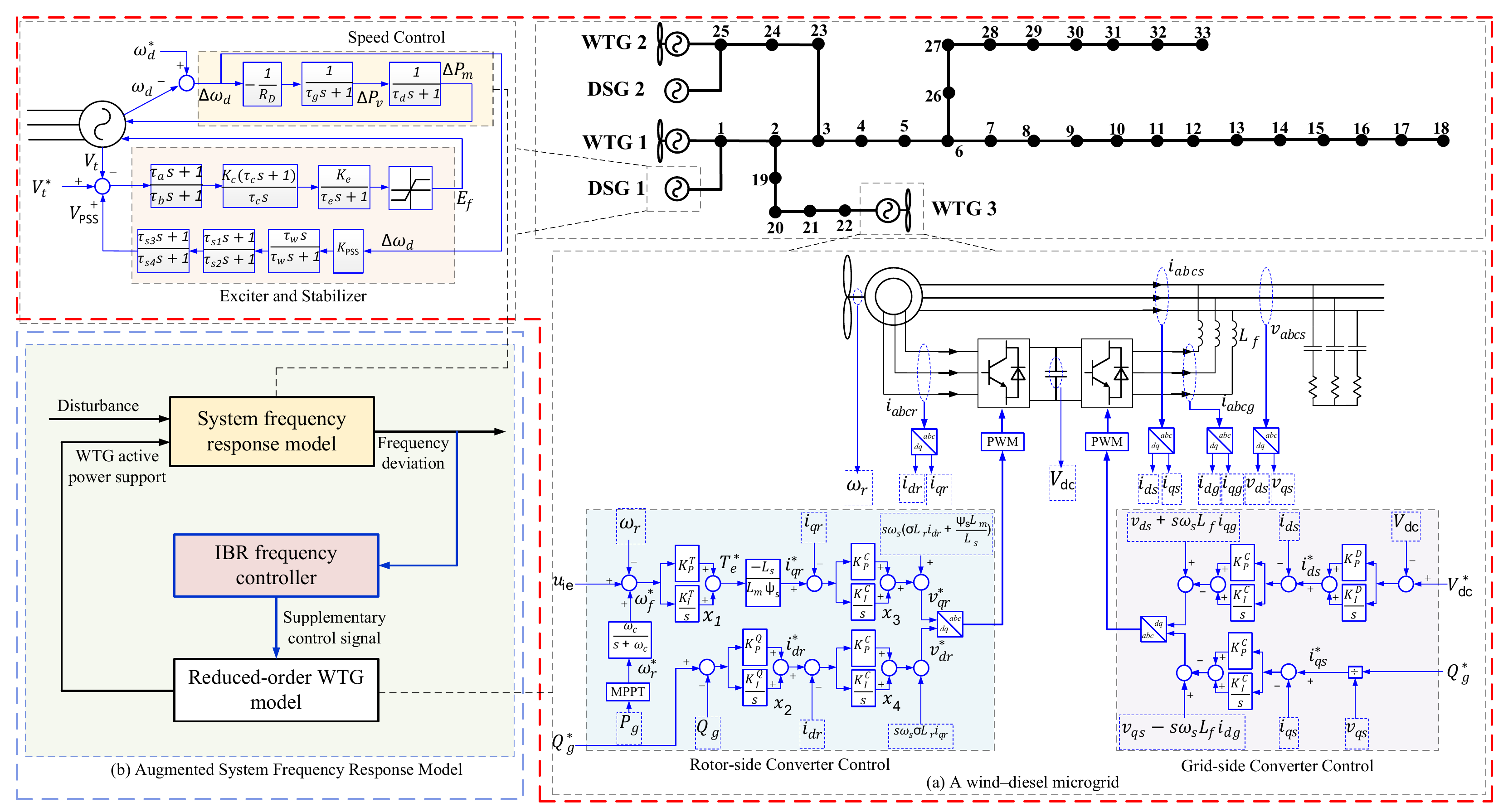}
	\caption{(a) Overall configuration of the diesel–wind microgrid. (b) Reduced-order augmented frequency response model.}
	\label{fig_microgrid}
\end{figure*}

The ASFR model enhances the conventional SFR model by adding an additional feedback loop that provides fast active power support from IBRs using system frequency as the input, as shown in  Fig.~\ref{fig_microgrid} (b). 

During a power imbalance disturbance, frequency deviations will be sensed by the IBR frequency controller to generate a supplementary control signal. Consequently, the IBR injects or absorbs active power to improve frequency regulation in low-inertia microgrids. It is well known that the conventional SFR model can dominantly capture the frequency behavior of synchronous generators \cite{kundur1994power, andersonLoworderSystemFrequency1990}. Hence, the key success of the ASFR is a reduced-order IBR model that captures its power variation under the supplementary control signal. There are several methods to complete this objective, including the selective modal analysis (SMA) reduction method \cite{hector,zhangProvisionGuaranteedInertial2018} and balanced truncation model reduction \cite{morovatiRobustOutputFeedback2021}. Detailed derivations are out of scope of this paper and can be referred to these references \cite{hector,zhangProvisionGuaranteedInertial2018,morovatiRobustOutputFeedback2021}. Let the reduced-order WTG model be represented as follows \cite{zhangProvisionGuaranteedInertial2018,zhangSetTheoryBasedSafety2019}:
\begin{subequations}
\begin{align}
\Delta \dot{\omega}_{r} &= A_{rd}\,\Delta \omega_{r} + B_{rd}\,u
\label{eq:wtg1}
\\
\Delta P_{g} &= C_{rd}\,\Delta \omega_{r} + D_{rd}\,u
\label{eq:wtg2}
\end{align}
\end{subequations}
where $\Delta \omega_r$ denotes the rotor speed deviation, $u$ represents the supplementary frequency support control signal, and $\Delta P_g$ denotes the active power signal injected by the WTG into the microgrid. The diesel synchronous generator frequency dynamics are governed by the SFR model augmented with the WTG support:
\begin{subequations}
\begin{align}
2H\,\Delta\dot{\omega}_d &= \Delta P_m - \Delta P_e + \Delta P_g\label{eq:swing}\\
T_d\,\Delta \dot{P}_{m} &= -\Delta P_{m} + \Delta P_{v}\label{eq:engine}\\
T_s\,\Delta \dot{P}_{v} &= -\Delta P_{v}- \frac{\Delta \omega_{d}}{R_{d}}\label{eq:governor}
\end{align}
\end{subequations}
where $H$ is the inertia constant, $T_d$ and $T_s$ are the time constants of the diesel engine and governor, respectively, $R_d$ is the governor droop, $\Delta\omega_d$ is the deviation of the generator speed, $\Delta P_m$ is the deviation of mechanical power, $\Delta P_v$ is the reference valve position from the governor, $\Delta P_e$ is the disturbance of electrical load, and $\Delta P_g$ is the injection of additional power from the wind turbine. Defining the state vector as $x = \begin{bmatrix}\Delta\omega_{d},\Delta P_{m},\Delta P_{v},\Delta \omega_{r}\end{bmatrix}^{\top}$, $u$ as the control input, and $\Delta P_e$ as the disturbance input, the ASFR model can be represented in compact state-space form as:
\begin{subequations}
\label{eq:state-space}
\begin{align}
A=\left[ \begin{array}{cccc} 
0 & \dfrac{1}{2H} & 0 & \dfrac{C_{rd}}{2H} \\[6pt]
0 & -\dfrac{1}{T_d} & \dfrac{1}{T_d} & 0 \\[6pt]
-\dfrac{1}{T_s R_d} & 0 & -\dfrac{1}{T_s} & 0 \\[6pt]
0 & 0 & 0 & A_{rd}
\end{array} \right]\\
B^{T}_1=\left[\begin{array}{cccc} \dfrac{D_{rd}}{2H} & 0 & 0 & B_{rd}  \end{array} \right] \\
B^{T}_2=\left[\begin{array}{cccc} -\dfrac{1}{2H} & 0 & 0 & 0  \end{array} \right] 
\end{align}
\end{subequations}
For the subsequent control design, the continuous-time model in (\ref{eq:state-space}) is discretized using a zero-order-hold approximation with sampling interval $T_s$. The resulting discrete-time representations are denoted as $A_d$, $B_{d1}$, $B_{d2}$, respectively.

\subsection{Offline Numerical Optimal Control}
To generate the optimal control signal for WTGs under different disturbances, an offline numerical control problem is formulated as follows \cite{zhangRobustTrajectoryconstrainedFrequency2023}:
\begin{equation}
\min_{u_i(k)} \; C_U = \sum_{i=1}^{N_w} \sum_{k=1}^{T} 
[u_i(k) + w_1 \Delta u_{1} + w_2 \Delta u_{2}]
\label{eq:objective}
\end{equation}
subject to:
\begin{subequations}
\begin{align}
x(k+1) &= A_d x(k) + B_{d1}u(k) + B_{d2}\Delta P^{s}_{e} \label{eq:dynamics} \\
\Delta\omega_d(k) &\leq \Delta \omega_{d,\text{lim}} \label{eq:freq_lim} \\
-u_{\text{lim}} &\leq u_i(k) \leq u_{\text{lim}} \label{eq:control_bound} \\
\Delta u_{2} &\leq u_i(k)-u_i(k-1) \leq \Delta u_{1}  \label{eq:limit_switching} \\
x(0) &= X_0, \quad u_i(0) = 0, \quad u_i(T) = 0 \label{eq:boundary}
\end{align}
\end{subequations}
Eq. \eqref{eq:dynamics} is the constraint of system dynamics under a sampled disturbance $\Delta P^{s}_{e}$. The frequency constraint~\eqref{eq:freq_lim} ensures diesel frequency deviation remains within safety limits. Together with the last two terms in the objective, constraint~\eqref{eq:limit_switching} limits the rate of change of the control signal. Constraint \eqref{eq:boundary} defines the initial and final conditions of states and control. The total control effort is minimized in \eqref{eq:objective}. This problem is a linear program and can be efficiently solved under a large number of disturbance samples. Hence, sufficient samples of optimal state-action trajectories can be generated for training the DNN controller.

\section{Lyapunov Stability Analysis of the Learned Frequency Controller} \label{sec_lyapunov}
In this section, we establish the closed-loop stability condition of the proposed DNN-based frequency controller. The method isolates the NN nonlinearities, characterizes their behavior through sector quadratic constraints, and obtains Lyapunov stability conditions using the absolute stability theory \cite{khalil2002nonlinear,hu2001control}. The resulting stability-constrained learning formulations are then solved using the alternating direction method of multipliers (ADMM) algorithm ~\cite{yinImitationLearningStability2022}.

\subsection{Neural Network Representation}
\begin{figure}[t]
    \centering
    \includegraphics[clip,width=0.6\linewidth,trim=0.5cm 0.5cm 0.5cm 0.5cm]{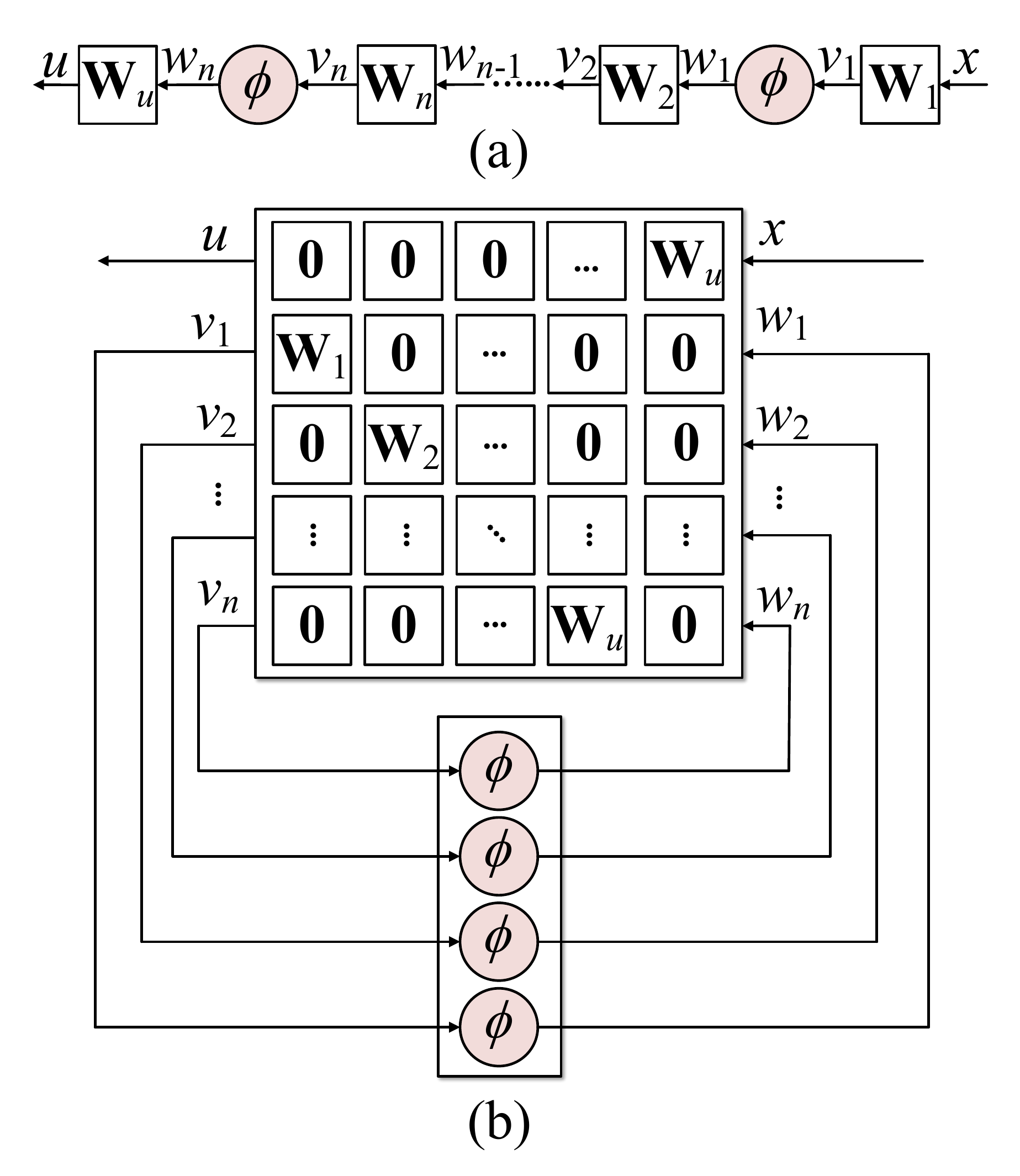}
    \vspace{-10pt}
    \caption{DNN restructuring to isolate nonlinear activation functions.}
    \vspace{-15pt}
    \label{fig_dnn_reorg}
\end{figure}

The usage of the absolute stability theory and sector quadratic constraints requires an isolation of nonlinear components \cite{hu2001control}, which are the activation functions in the DNN-based frequency controller. To decouple the nonlinear activation functions from the linear weights~\cite{kimStandardRepresentationUnified2018}, the DNN controller $u = \pi(x;\theta)$ is rewritten in the compact form \cite{yinStabilityAnalysisUsing2022}:
\begin{equation} \begin{bmatrix} u(k) \\ v(k) \end{bmatrix} = N \begin{bmatrix} x(k) \\ w(k) \end{bmatrix}, \quad w(k) = \phi(v(k)) 
\label{eq:nn_rep} 
\end{equation}
where $v_\phi$ and $w_\phi$ denote the stacked pre- and post-activation signals, respectively. The matrix $N$ is partitioned as:
\begin{equation}
N = \begin{bmatrix} N_{ux} & N_{uw} \\ N_{vx} & N_{vw} \end{bmatrix}
\label{eq:N_partition}
\end{equation}
and $\phi(\cdot)$ is applied element-wise. Fig. \ref{fig_dnn_reorg} illustrates this restructuring, including the structure of matrix $N$. Bias parameters are not considered. Expanding \eqref{eq:nn_rep} and \eqref{eq:N_partition} yields a representation of $u$ in terms of states and stacked pre-activation signals:
\begin{subequations}
\begin{align}
&u(k) = N_{ux}x(k) + N_{uw}w_{\phi}(k)\label{eq:nn_expand_1}\\
&v_{\phi}(k) = N_{vx}x(k) + N_{vw}w_{\phi}(k)\label{eq:nn_expand_2}\\
&w_{\phi}(k) =\phi(v_{\phi}(k))= \phi(N_{vx}x(k) + N_{vw}w_{\phi}(k))\label{eq:nn_expand_3}
\end{align}
\label{eq:nn_expand}
\end{subequations}

\subsection{Sector Quadratic Constraints and Lyapunov Condition}
To analyze the closed-loop behavior of the DNN controller, the activation functions must first be characterized. Since tanh and ReLU are bounded nonlinearities, they satisfy well-known sector conditions~\cite{wanAdapSafeAdaptiveSafeCertified2023, fazlyabSafetyVerificationRobustness2022}, meaning their outputs stay within a predictable range for any given input. Based on this, we assume $\phi$ satisfies a local sector $[\alpha_\phi, \beta_\phi]$ element-wise, where $\alpha_\phi,\beta_\phi\in\mathbb{R}^{\phi_n}$. With these bounds in hand, the pre- and post-activation signals $v_\phi$ and $w_\phi$ satisfy \cite{yinImitationLearningStability2022}:
\begin{equation}
\begin{bmatrix} v_\phi \\ w_\phi \end{bmatrix}^\top
\begin{bmatrix} 
-2A_\phi B_\phi \Lambda & (A_\phi+B_\phi)\Lambda \\ 
(A_\phi+B_\phi)\Lambda & -2\Lambda 
\end{bmatrix}
\begin{bmatrix} v_\phi \\ w_\phi \end{bmatrix} \geq 0
\label{eq:qc}
\end{equation}
where $A_\phi = \mathrm{diag}(\alpha_\phi)$ and $B_\phi = \mathrm{diag}(\beta_\phi)$ are the lower and upper sector bound matrices, and $\Lambda = \mathrm{diag}(\lambda) \succeq 0$ is a diagonal scaling matrix~\cite{megretskiSystemAnalysisIntegral1997}. Substituting \eqref{eq:nn_expand_2} into \eqref{eq:qc} yields the sector condition represented by $x$ and $w$:
\begin{align}
	\begin{aligned}
		\left[\begin{array}{c}
			x\\ w
		\end{array}
		\right]^{T}
		\left[\begin{matrix}
		N_{vx} & N_{vw}\\
			0 & I \end{matrix}
        \right]^{T}
		\left[\begin{matrix}
			- 2 A_{\phi} B_{\phi}  & \left(A_{\phi} + B_{\phi}\right) \\
			\left(A_{\phi} + B_{\phi} \right)  & - 2 \end{matrix}
        \right]\\
		\left[\begin{matrix}
			N_{vx} & N_{vw}\\
			0 & I \end{matrix}\right]
		\left[\begin{array}{c}
			x\\ w
		\end{array}
		\right]\succ 0 
	\end{aligned}
    \label{eq:qc_xw}
\end{align}

To guarantee closed-loop stability, we define a quadratic Lyapunov function $V(x) = x^\top Px$ with $P \succ 0$~\cite{ boyd1994linear}. Closed-loop stability requires that $V(x)$ must strictly decrease at every step:
\begin{equation}
V(x(k+1)) - V(x(k)) < 0, \quad \forall\, x(k) \neq 0
\label{eq:lyap_decrease}
\end{equation}
Substituting $V(x) = x^\top Px$ into~\eqref{eq:lyap_decrease} gives:
\begin{equation}
(Ax + Bu)^\top P 
(Ax + Bu) - x^\top Px < 0
\label{eq:lyap_cond}
\end{equation}
Note that we drop the time step notation $k$ since they are all in the same time step, and \eqref{eq:lyap_cond} will hold for all $k$. Substituting \eqref{eq:nn_expand_1} and \eqref{eq:nn_expand_3} into \eqref{eq:lyap_cond} yields the closed-loop stability condition:
\begin{align}
	\begin{aligned}
		\left[\begin{array}{c}
			x\\ w
		\end{array}
		\right]^{T}
		\left[\begin{array}{cc}
			I & N_{ux}^{T}\\
			0 & N_{uw}^{T}
		\end{array}
		\right]
		\left[\begin{array}{cc}
			A^{T} P A  - P  & A^{T} P B  \\ 
			B^{T} P A  &  B^{T} P B 
		\end{array}
		\right]\\
		\left[\begin{array}{cc}
			I & 0\\ 
			N_{ux} &  N_{uw}
		\end{array}
		\right]
		\left[\begin{array}{c}
			x\\ w
		\end{array}
		\right]\prec 0 
	\end{aligned}
    \label{eq:lyap_closed}
\end{align}
Once the activation function is determined, proper sector condition $[\alpha_\phi, \beta_\phi]$ can be selected to ensure the satisfaction of \eqref{eq:qc} and therefore \eqref{eq:qc_xw}. Whenever \eqref{eq:qc_xw} holds, we like to train proper DNN, whose parameters are represented by $N$, so that \eqref{eq:lyap_closed} holds. Note that both \eqref{eq:qc_xw} and \eqref{eq:lyap_closed} will hold for all $x$ and $w_\phi$. We can apply the S-procedure for \eqref{eq:qc_xw}$\Rightarrow$\eqref{eq:lyap_closed}, which yields the closed-loop stability condition in the form of a matrix inequality:
\begin{align}
\begin{aligned}
&
\begin{bmatrix}
I & N_{ux}^{T} \\
0 & N_{uw}^{T}
\end{bmatrix}
\begin{bmatrix}
A^{T} P A - P & A^{T} P B \\
B^{T} P A & B^{T} P B
\end{bmatrix}
\begin{bmatrix}
I & 0 \\
N_{ux} & N_{uw}
\end{bmatrix}
\\[1ex]
&
+ \lambda
\begin{bmatrix}
N_{wx} & N_{wu} \\
0 & I
\end{bmatrix}^{T}
\begin{bmatrix}
-2 A_{\phi} B_{\phi} & A_{\phi} + B_{\phi} \\
A_{\phi} + B_{\phi} & -2 I
\end{bmatrix}
\begin{bmatrix}
N_{wx} & N_{wu} \\
0 & I
\end{bmatrix}
\prec 0 .
\end{aligned}
\label{eq:lyap_closed_MI}
\end{align}

\subsection{Convex LMI Conditions}
The stability condition~\eqref{eq:lyap_closed_MI} is nonconvex since both the network weights $N$ and the Lyapunov matrix $P$ are decision variables. First, a loop transformation to normalize $\phi$ to the sector $[-1,1]$ is applied \cite{yinImitationLearningStability2022}, which will lead to the following conversion in \eqref{eq:lyap_closed_MI}:
\begin{align}
	\begin{aligned}
        \left[\begin{matrix}
			- 2 A_{\phi} B_{\phi} \Lambda & \left(A_{\phi} + B_{\phi} \right) \Lambda\\
			\left(A_{\phi} + B_{\phi} \right) \Lambda & - 2 \Lambda\end{matrix}
            \right] \Rightarrow
            \left[\begin{matrix}
			2 \Lambda & 0\\
			0 & - 2 \Lambda\end{matrix}
            \right] 
	\end{aligned}
\end{align}
and
\begin{subequations}
\begin{align}
& \tilde{N}_{ux}=N_{ux} + N_{uw}\Gamma(I-N_{vw}\Gamma)^{-1}N_{vx}\\
& \tilde{N}_{uz}=N_{uw}\Delta + N_{uw}\Gamma(I-N_{vw}\Gamma)^{-1}N_{vw}\Delta\\
& \tilde{N}_{vx}=(I-N_{vw}\Gamma)^{-1}N_{vx}\\
& \tilde{N}_{vz}=(I-N_{vw}\Gamma)^{-1}N_{vw}\Delta
\end{align}
\label{eq:loop_shaping}
\end{subequations}
where
\begin{align}
& \Gamma=\frac{A_\phi+B_\phi}{2},\Delta=\frac{B_\phi-A_\phi}{2}
\end{align}
The loop shaping technique will transform \eqref{eq:lyap_closed_MI} into:
\begin{align}
	\begin{aligned}
		\left[\begin{matrix}
			A^{T} + \tilde{N}_{ux}^{T} B^{T} & \tilde{N}_{vx}^{T} \\
			\tilde{N}_{uz}^{T} B^{T} & \tilde{N}_{vz}^{T}
		\end{matrix}\right]
		\left[\begin{matrix}
			P & 0\\
			0 & \Lambda
		\end{matrix}\right]
		\left[\begin{matrix}
			A + B \tilde{N}_{ux} & B \tilde{N}_{uz}  \\
			\tilde{N}_{vx} & \tilde{N}_{vz}
		\end{matrix}\right]\\
		-\left[\begin{matrix}
			P & 0\\
			0 & \Lambda
		\end{matrix}\right] \prec 0
	\end{aligned}
    \label{eq:lyap_closed_loop}
\end{align}
Introduce the following change of variables:
\begin{subequations}
\begin{align}
& Q_1 = P^{-1}\succ 0,Q_2 = \Lambda^{-1}\succ 0\label{eq:change_var_1}\\
&\tilde{N}Q=\left[\begin{matrix}
			\tilde{N}_{ux}Q_1 & \tilde{N}_{uz}Q_2 \\
			\tilde{N}_{vx}Q_1 & \tilde{N}_{vz}Q_2
		\end{matrix}\right]=
		\left[\begin{matrix}
			L_1 & L_2\\
			L_3 & L_4
		\end{matrix}\right]=L\label{eq:change_var_2}
\end{align}
\label{eq:change_var}
\end{subequations}
where
\begin{align}
        Q=\left[\begin{matrix}
			Q_1 & 0\\
			0 & Q_2
		\end{matrix}\right]
\end{align}
Using Schur complements with the change of variables in \eqref{eq:change_var} will lead to the LMI representing the stability under sector condition:
\begin{equation}
\begin{bmatrix} 
Q_1 & 0 & Q_1A^\top + L_1^\top B^\top & L_3^\top \\ 
0 & Q_2 & L_2^\top B^\top & L_4^\top \\ 
AQ_1 + BL_1 & BL_2 & Q_1 & 0 \\ 
L_3 & L_4 & 0 & Q_2 
\end{bmatrix} \succ 0
\label{eq:lmi_stability}
\end{equation}
This condition~\eqref{eq:lmi_stability} is convex in new decision variables $Q_1, Q_2, L_1,\ldots,L_4$ and can be solved efficiently using standard semi-definite programming (SDP) solvers ~\cite{boyd1994linear}. Constraint~\eqref{eq:lmi_stability} guarantees that the closed-loop stability of the system \eqref{eq:dynamics} under the DNN feedback \eqref{eq:nn_expand} as long as the DNN's activation functions satisfy the sector conditions by proper selection of $A_\phi$ and $B_\phi$.

\subsection{Stability-Constrained Training via ADMM}\label{sec_ADMM}
To embed the LMI stability conditions directly into the DNN training, we solve the following constrained imitation learning problem
\begin{subequations}
\begin{align}
\min_{N,Q,L} \; \eta_1\mathcal{L}(N) - 
\eta_2\log\det(Q_1) \quad \\
\text{subjected to }\text{ constraints }\eqref{eq:lmi_stability}\text{,  }\eqref{eq:change_var}\text{, and }\eqref{eq:loop_shaping}
\end{align}\label{eq:admm_prob}
\end{subequations}
where $\mathcal{L}(N)$ measures how closely the DNN output matches the expert control actions, and the term $-\eta_2\log\det(Q_1)$ the conservatism of the stability condition. The weights $\eta_1$ and $\eta_2$ balance between imitation accuracy and conservatism. 
Directly solving \eqref{eq:admm_prob} with a random initialization of $N$ will not satisfy \eqref{eq:lmi_stability} and \eqref{eq:change_var_2} simultaneously. The general idea is to relax constraint \eqref{eq:change_var_2} and minimize $\tilde{N}Q-L$ iteratively.
To handle this, we define the following augmented loss function~\cite{yinImitationLearningStability2022}:
\begin{subequations}
\begin{align}
\mathcal{L}_a(N,Q,L,Y)=\eta_1\mathcal{L}(N) - 
\eta_2\log\det(Q_1)\\+\text{tr}[Y^T\bullet(\tilde{N}Q-L)]+\rho/2||\tilde{N}Q-L||^{2}_{F}
\end{align}
\end{subequations}
where $Y$ is the Lagrange multiplier, $||\bullet||^{2}_{F}$ is the Frobenius norm, and $\rho>0$ is the regularization parameter. The ADMM proceeds as follows:
\begin{itemize} 
\item \text{$N$-update:} The network weights $N$ are updated by minimizing the augmented Lagrangian via gradient descent together with constraint \eqref{eq:change_var_2}, keeping $Q$, $L$, and $Y$ fixed.
\item \text{($Q$, $L$)-update:} With $N$ and $Y$ fixed, the augmented Lagrangian is minimized together with constraints \eqref{eq:lmi_stability} and \eqref{eq:change_var_2} to update $Q$ and $L$. 
\item \text{$Y$-update:} The dual variable $Y$ is updated using the relaxed constraint residual $\tilde{N}Q - L$. If $||\tilde{N}Q-L||_{F}$ is less than the stopping tolerance, then the algorithm is considered to converge.
\end{itemize}

\section{Case Study Between Linear and DNN Feedback}
\subsection{Linear Feedback Control Baseline }
To assess the necessity of DNN feedback, a traditional linear feedback control policy is first considered in imitation learning. The controller employs a static state-feedback structure of the form $u = Kx$. The controller gains are obtained by solving an optimization problem subject to the Kalman constraint \cite{palanFittingLinearControl2020}:
\begin{subequations}
\begin{align}
&\min_{K,P} \; \mathcal{L}(K) + r(K)\\
&\text{subject to }Q+A^TP(A+BK)-P=0\\
& RK+B^TP(A+BK)=0\\
& P \succeq 0, \quad Q \succeq 0, \quad R\succeq I
\end{align}\label{eq:IL_linear}
\end{subequations}
where $\mathcal{L}(K)$ represents the loss function, while $r(K)$ penalizes excessively large feedback gains. This problem can be solved using a similar approach discussed in Section \ref{sec_ADMM}.

\begin{figure}[h!]
    \centering
    \includegraphics[width=0.8\columnwidth]{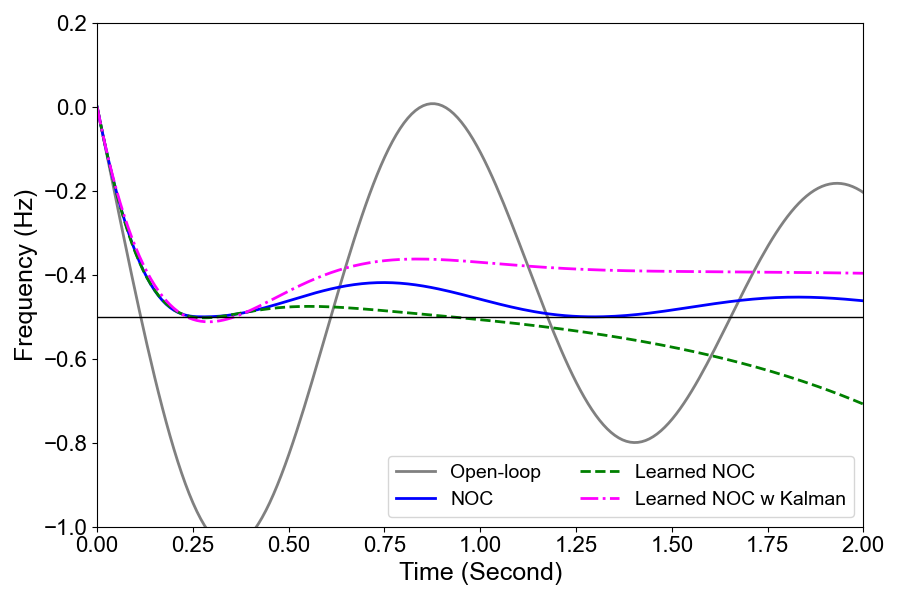}
    \caption{Frequency performance under open-loop condition, offline NOC, learned NOC feedback without and with the Kalman constraints.}
    \label{fig:frequency_response}
\end{figure}
First, a pair of state and expert control data is generated under a fixed disturbance. Then, problem \eqref{eq:IL_linear} is trained using these data. Fig.~\ref{fig:frequency_response} illustrates the frequency performance under open-loop condition, offline NOC, learned NOC feedback without and with the Kalman constraints, respectively. As we can see, the frequency responses under learned NOC feedback can meet the nadir constraint regardless of the Kalman constraints, demonstrating the effectiveness of imitation learning. However, the frequency trajectory under learned NOC feedback without Kalman constraints is unstable as it begins to diverge after reaching the first nadir. In contrast, incorporating the Kalman constraints results in a stable and safe frequency trajectory, as shown by the dashed magenta curve.

\begin{figure}[h!]
    \centering
    \includegraphics[width=0.9\columnwidth]{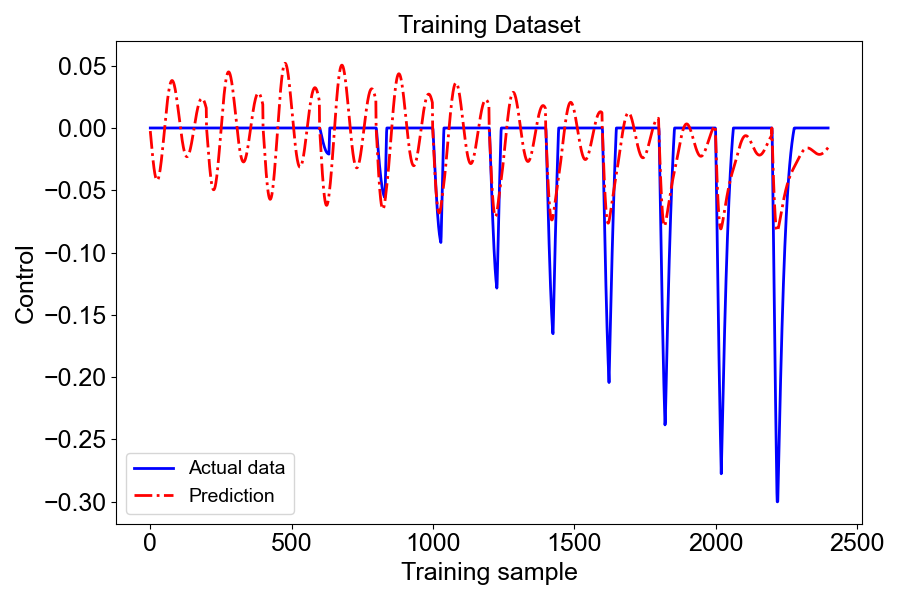}
    \caption{Comparison between expert control signals and linear policy predictions in the training dataset.}
    \label{fig:training_dataset}
\end{figure}

Second, multiple sets of state and expert control data are generated under varying disturbances. The expert control actions are shown in Fig. \ref{fig:training_dataset}, where the corresponding state-control pairs are used to train problem \eqref{eq:IL_linear}. The predicted control signals are shown as dashed red curves in Fig. \ref{fig:training_dataset}. As we can see, linear control is unable to mimic the expert control under varying disturbance magnitudes due to its representation limitation.

\subsection{Stability-Constrained DNN Controller Performance}
\begin{figure}[h!]
    \centering
    \includegraphics[width=0.9\columnwidth]{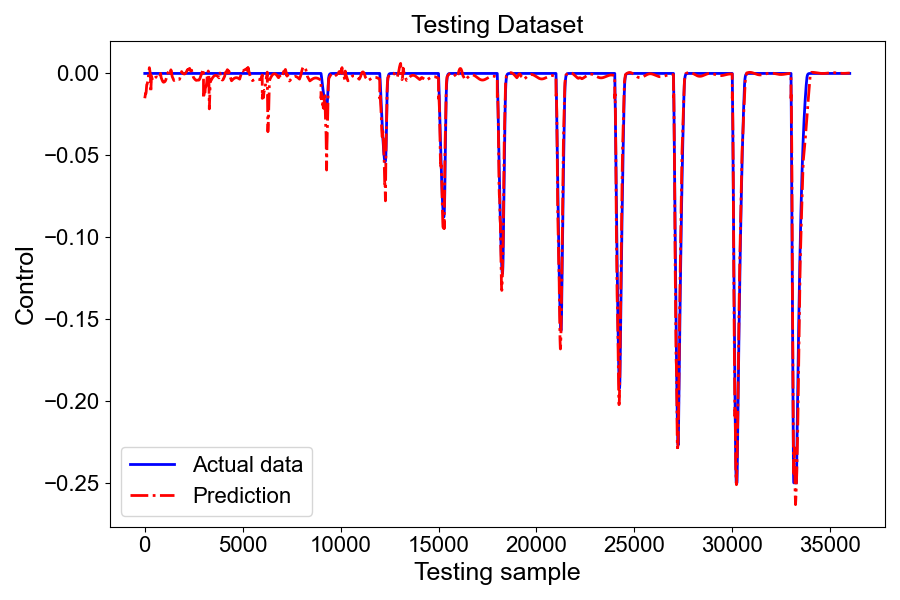}
    \caption{Comparison between expert control signals and DNN predictions on the testing dataset.}
    \label{fig:testing_dataset}
\end{figure}

The same sets of state and expert control data are used to train problem \eqref{eq:admm_prob} to obtain the DNN controller. Its prediction accuracy is evaluated using an independent testing dataset, as illustrated in Fig.~\ref{fig:testing_dataset}. The predicted signals closely follow the expert  control actions under different disturbance scenarios, which indicates that the DNN can accurately reproduce the expert control behavior.

\begin{figure}[h!]
    \centering
    \includegraphics[width=0.9\columnwidth]{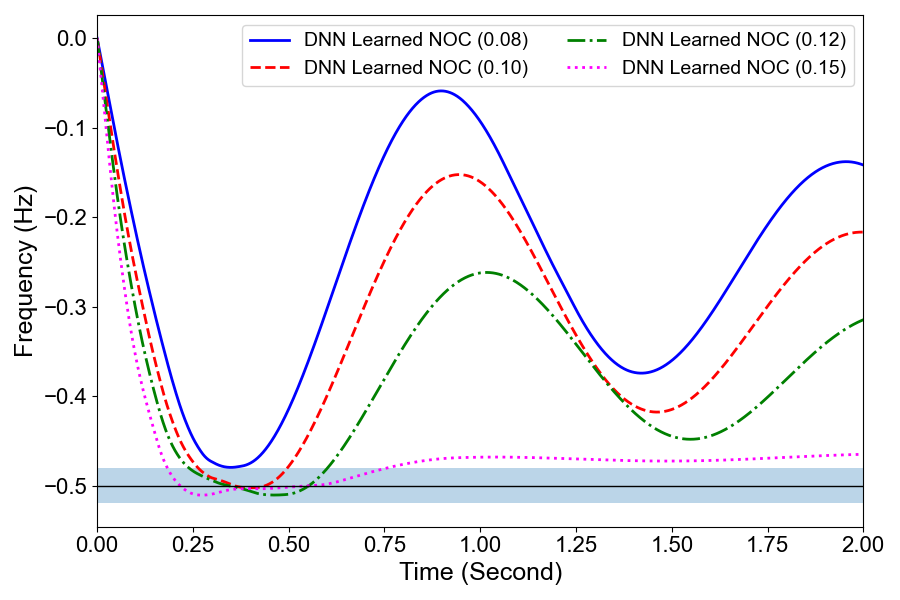}
    \caption{Closed-loop frequency response obtained using the stability-constrained DNN controller under different disturbances.}
    \label{fig:dnn_frequency}
\end{figure}

The effectiveness of the learned DNN controller on the ASFR model in \eqref{eq:dynamics} is shown in Fig.~\ref{fig:dnn_frequency}. As shown, the DNN controller effectively limits frequency deviation under different disturbances. On the other hand, the DNN controller is trying to minimize the control effort of the WTG as every frequency trajectory just resides around the nadir limit. Due to the prediction errors, some frequency trajectories slightly violate the nadir constraint. It is worth noting that such transient frequency violations typically do not trigger underfrequency load shedding relays because the frequency trajectory does not remain within the relay trip region long enough to satisfy the relay's time-delay characteristic. In addition, we can set the safety limit more conservative to count for the prediction errors. 

\section{Case Study: 33-Node Based Microgrid} \label{sec_case_33}
To validate the robustness and scalability of the proposed decentralized DNN-based control framework, comprehensive simulations are conducted on the modified IEEE 33-node microgrid test system implemented in MATLAB/Simulink. The test system comprises two diesel synchronous generators (DSG 1 and 2) connected at bus 1 and bus 25, respectively, and three WTGs connected at bus 1, bus 25, and bus 22, respectively, as illustrated in Fig. \ref{fig_microgrid}.

The intelligence built in DNN naturally admits a decentralized implementation when the control is designed for multiple DSG-WTG pairs. Each DNN controller is associated with a local DSG-WTG pair and utilizes only locally available measurements required for frequency regulation, including frequency deviation $(\Delta\omega_d)$, governor states $(\Delta P_m,\Delta P_v)$, and rotor speed deviation $(\Delta\omega_r)$.  WTGs do not need to communicate about their supportive power since the DNN control will implicitly convert the trajectory information into net power imbalance in real time. Such a \emph{decentralized intelligent feature} improves scalability, reduces communication dependency, and enhances the practical feasibility of large-scale microgrids with multiple IBRs.

\vspace{-10pt}
\begin{table}[h!]
\caption{33-Node Microgrid Generator and Controller Parameters}
\label{tab:mg_params}
\centering
\renewcommand{\arraystretch}{1}
\begin{tabular}{>{\raggedright\arraybackslash}p{2cm} >{\raggedright\arraybackslash}p{2cm} >{\raggedright\arraybackslash}p{3.2cm}}
\toprule
\textbf{Component} & \textbf{Parameter} & \textbf{Value} \\
\midrule
Diesel synchronous generators (DSG $i$, $i=1,2$)
& Rated power & $2~\mathrm{MW}$ \\
& Inertia constant & $H_{D,i}=1~\mathrm{s}$ \\
& Turbine time constant & $T_{d,i}=0.2~\mathrm{s}$ \\
& Governor time constant & $T_{s,i}=0.1~\mathrm{s}$ \\
\midrule
Wind turbine generators (WTG $i$, $i=1,2,3$)
& Rated power & $1~\mathrm{MW}$ \\
& Inertia constant & $H_{T,i}=4~\mathrm{s}$ \\
& Torque regulator & $K^{T}_{P,i}=2,\;\;K^{T}_{I,i}=0.1$ \\
& Reactive power regulator & $K^{Q}_{P,i}=1,\;\;K^{Q}_{I,i}=5$ \\
& Current loop regulator & $K^{C}_{P,i}=0.6,\;\;K^{C}_{I,i}=8$ \\
& DC-link voltage regulator & $K^{D}_{P,i}=1,\;\;K^{D}_{I,i}=400$ \\
\bottomrule
\end{tabular}
\end{table}

The parameters considered for the microgrid simulation are summarized in Table \ref{tab:mg_params}. The simulation is conducted using full-order nonlinear electromagnetic models. Averaged models are adopted for converters in WTGs. The lower frequency safety limit is set to 59.5 Hz.

A deadband is applied to the DNN control signal before the signal is sent to the torque control loop of the WTG. The purpose of this deadband is to suppress small prediction errors when the state deviations and actual control signals are zero, as shown in Fig. \ref{fig:testing_dataset}, because we cannot force DNN to output exact zero under no disturbance scenarios. Hence, a sufficiently small DNN control signal indicates the absence of a disturbance and, consequently, is forced to zero using the deadband. Continuously passing these prediction fluctuations under normal conditions could reduce the energy extraction efficiency of the WTG.

\subsection{Single DNN Feedback}
\begin{figure}[h!]
    \centering
    \includegraphics[width=0.8\columnwidth]{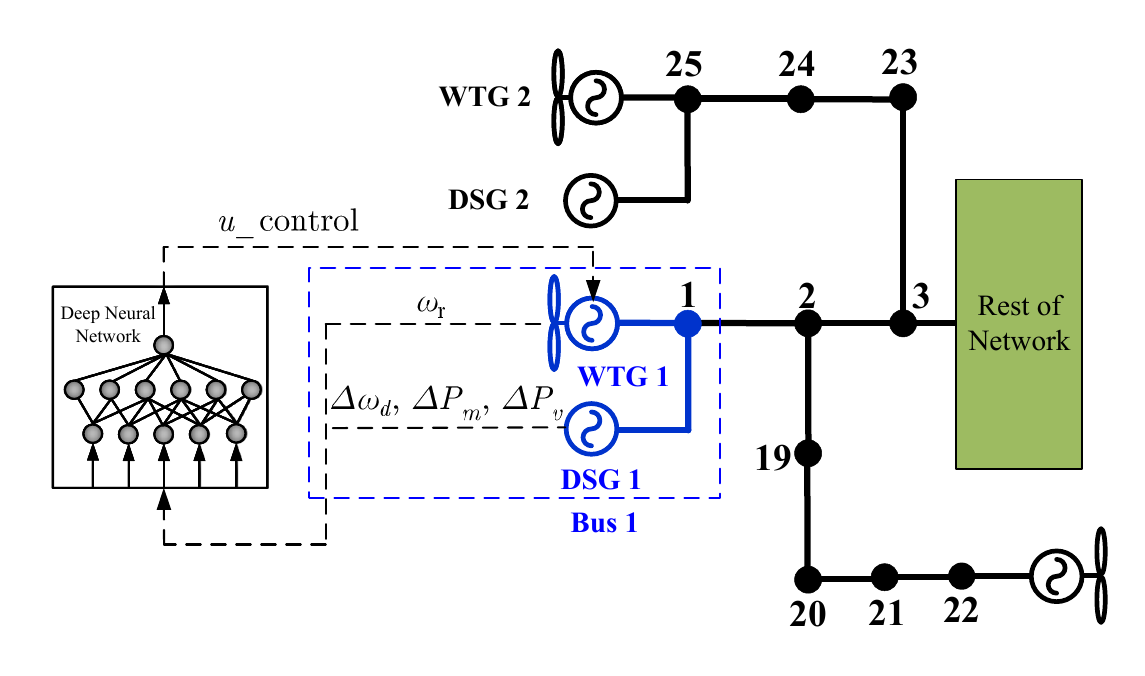}
    \caption{Control configuration at bus 1 with DNN controlled DSG-WTG.}
	\label{fig:bus1_control}
\end{figure}

For this study, the decentralized control configuration shown in Fig. \ref{fig:bus1_control} is considered. At bus 1, WTG 1 and DSG 1 are coordinated through the proposed DNN-based feedback framework, which utilizes locally measured frequency related states $(\Delta \omega_d,\Delta P_m,\Delta P_v,\Delta \omega_r)$ to generate the supplementary active power support signal for the WTG rotor-side converter. Under disturbance conditions, the WTG provides rapid active power injection, while the DSG contributes inertial response and slower governor response. In this configuration, the frequency response provided by DSG 2 does not need to be explicitly communicated to the DNN controller of DSG-WTG 1 to ensure the desired frequency performance. Instead, the frequency response of DSG 2 is reflected in the measured trajectories of DSG-WTG 1. The DNN implicitly captures this effect through the measured trajectories and maps them directly to the required control action in an end-to-end manner, without explicitly estimating the net power imbalance.

\begin{figure}[h!]
    \centering
    \includegraphics[width=0.9\columnwidth]{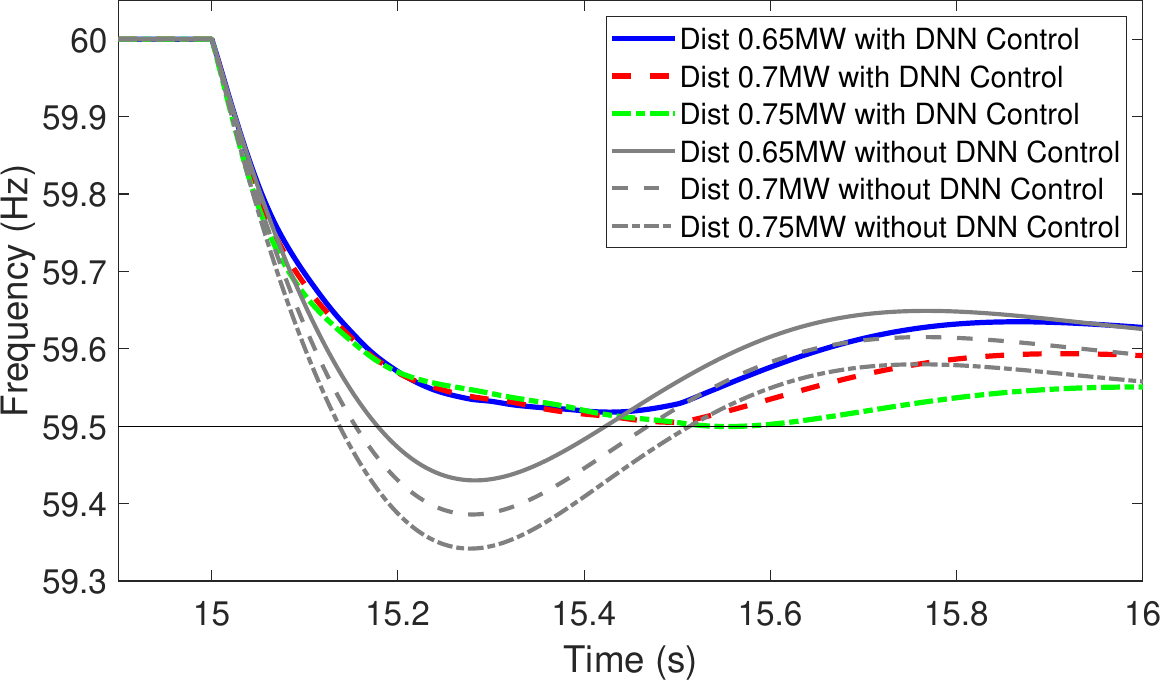}
    \caption{Nonlinear frequency response with and without DNN control under power imbalance disturbances.}
    \label{fig:freq_response}
\end{figure}

Fig. \ref{fig:freq_response} presents the frequency responses of DSG 1 under power imbalance disturbances of 0.65 MW, 0.70 MW, and 0.75 MW, both with and without DNN control. Without DNN control, the frequency nadir drops to approximately 59.35 Hz under the 0.75 MW disturbance, violating the prescribed safety threshold of 59.5 Hz, with increasingly deeper nadirs as the disturbance magnitude increases. This behavior indicates inadequate inertial and primary frequency response to support transient frequency safety when relying only on two DSGs. When the DNN controller is enabled, the frequency nadir is limited to approximately 59.5 Hz across all disturbance levels, maintaining system frequency within the prescribed safety threshold. The frequency performance guarantees of DSG 1 are achieved without measuring the states of DSG 2, verifying the decentralized intelligent feature. Furthermore, the DNN controllers provide near-optimal control signals in all three cases, as the frequency responses are not excessively enhanced but instead remain close to the prescribed safety limit.

\begin{figure}[h!]
    \centering
    \includegraphics[width=0.9\columnwidth]{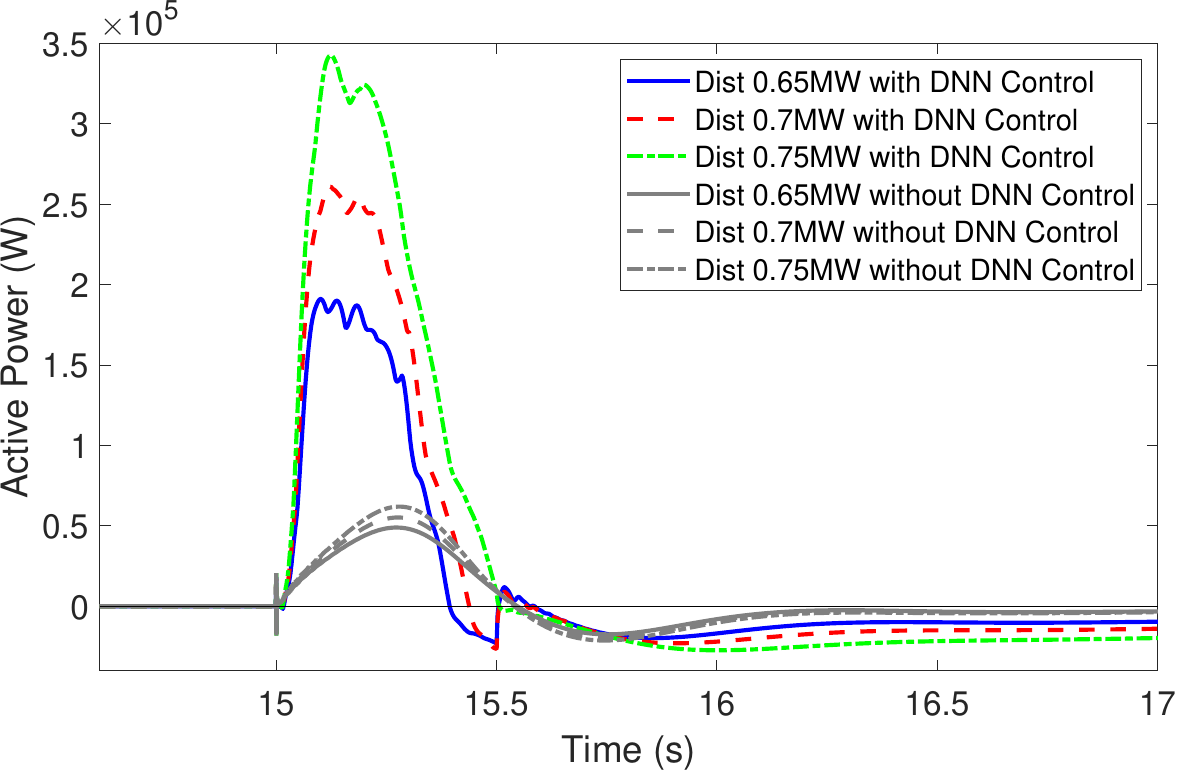}
    \caption{Active power injection of WTG 1 with and without DNN control under power imbalance disturbances.}
    \label{fig:Active_Power}
\end{figure}

The corresponding active power injections of WTG 1 under different cases are shown in Fig. \ref{fig:Active_Power}. Larger disturbances result in larger yet near-optimal power injections, demonstrating adaptive and optimized support capability. The fast active power injection characteristics directly contribute to the improved frequency responses observed in Fig. \ref{fig:freq_response}. Since the type-3 WTG is employed, there are small active power variations without DNN control due to the stator connection, which are small in magnitude and cannot effectively support the frequency. The jump at 15.5 seconds is due to the control deadband.
\vspace{-10pt}
\vspace{-10pt}
\vspace{-20pt}

\subsection{Two DNN Feedback}
\begin{figure}[h!]
    \centering
    \includegraphics[width=0.8\columnwidth]{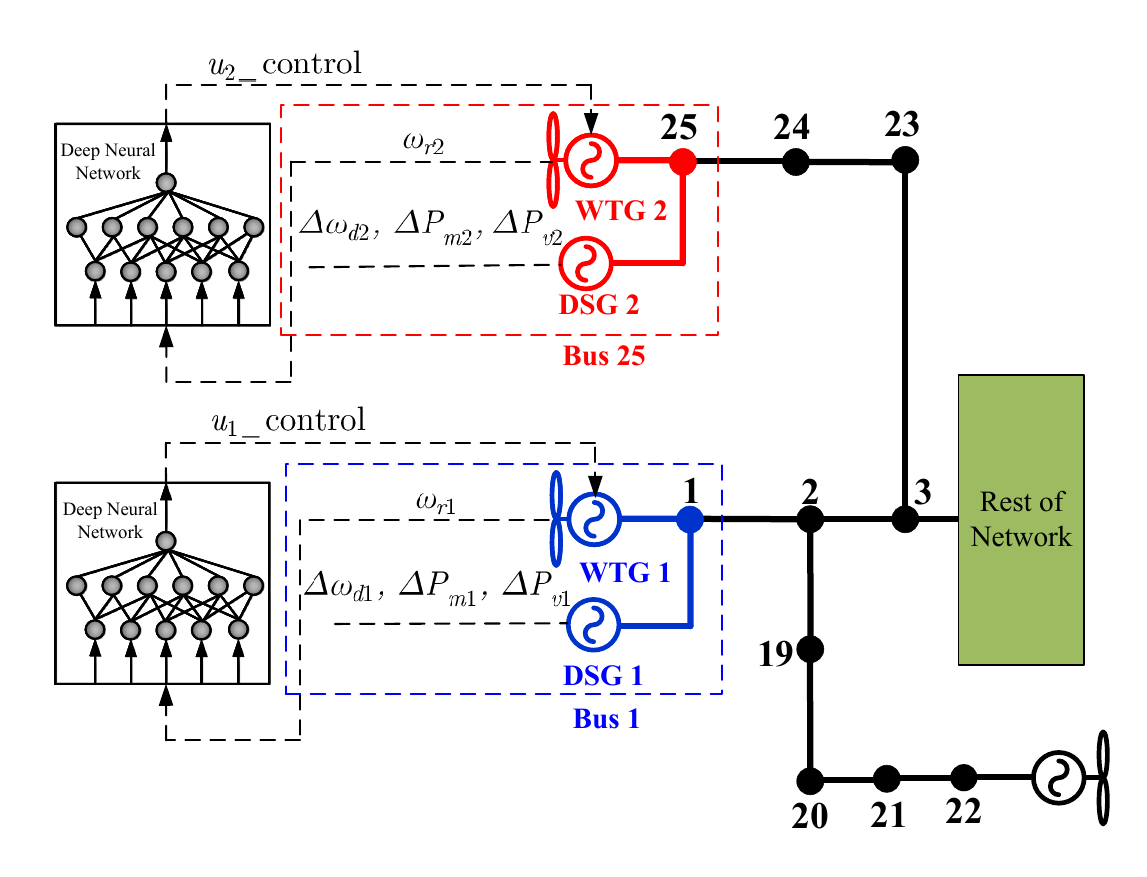}
    \caption{Control configuration at bus 1 and bus 25 with coordinated DSG-WTG.}
    \label{fig:buses_control}
\end{figure}

To further evaluate the scalability, a second DNN controller is applied at bus 25, as shown in Fig. \ref{fig:buses_control}. In this configuration, the DSG-WTG 1 pair at bus 1 and the DSG-WTG 2 pair at bus 25 operate as two decentralized local control units. Each controller independently generates supplementary frequency support signals using locally available measurements.

\begin{figure}[!t]
    \centering
    \includegraphics[width=0.5\textwidth]{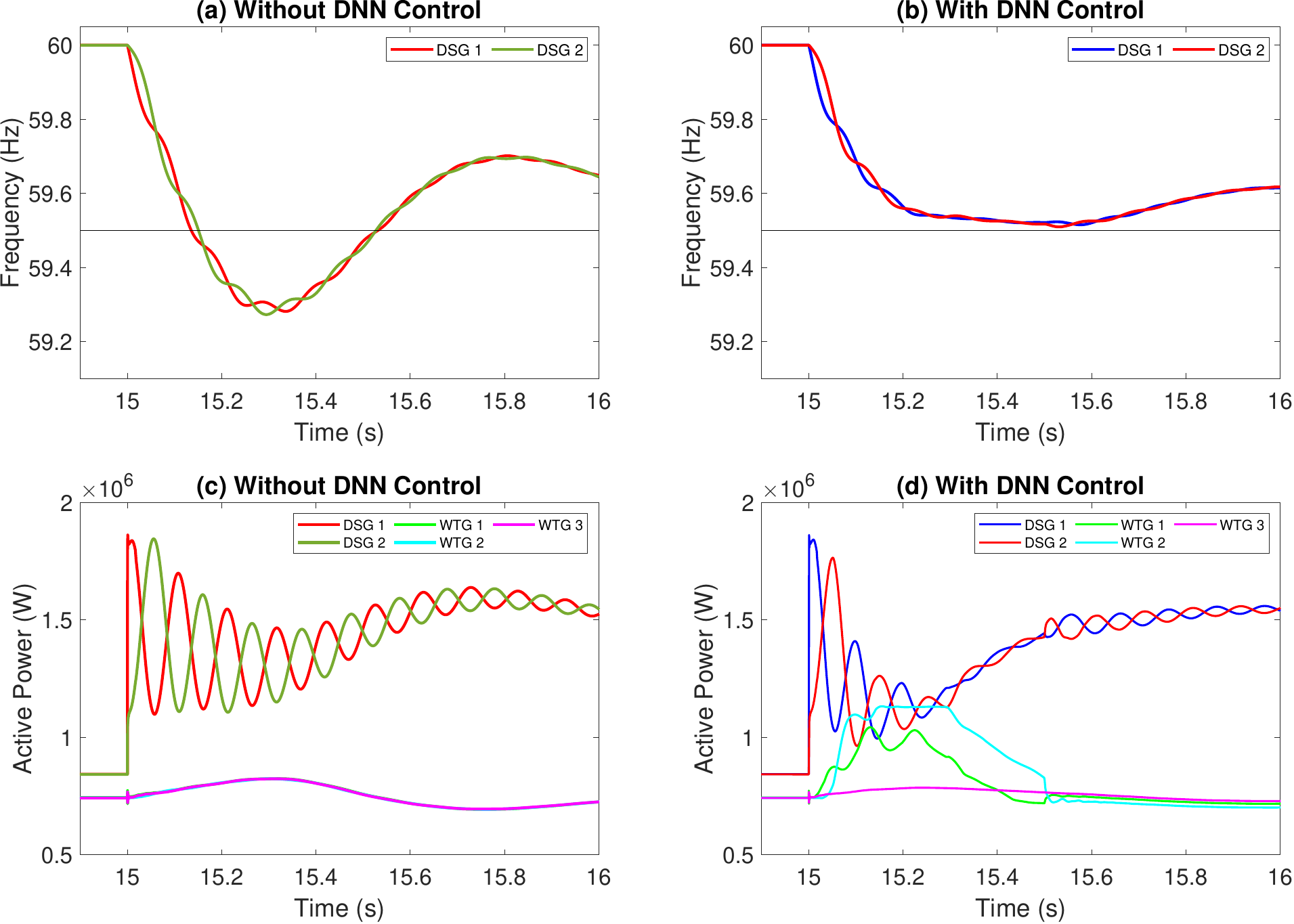}
    \caption{Dynamic frequency and active power responses of the microgrid system with and without DNN-based control.}
    \label{fig:dnn_nonlinear_response}
\end{figure}
\vspace{-0pt}

Fig. \ref{fig:dnn_nonlinear_response} presents the nonlinear frequency and active power responses under the multi-DNN configuration. As shown in Fig. \ref{fig:dnn_nonlinear_response} (b), the proposed control structure maintains the frequencies of both DSGs within the prescribed safety band under all disturbance conditions, while both DSGs' frequencies go beyond the limit without WTG's support, as shown in Fig. \ref{fig:dnn_nonlinear_response} (a). In addition, Fig. \ref{fig:dnn_nonlinear_response} (d) shows that the WTGs provide fast active power injections that reduce the inertial responses of DSGs, when compared with Fig. \ref{fig:dnn_nonlinear_response} (c), and therefore reduce the frequency declines. Both decentralized controllers can maintain frequency safety of individual DSGs without communication due to the decentralized intelligent feature discussed before.

This decentralized intelligence has a profound meaning. In the past, most performance guaranteed control paradigms require a centralized configuration, which is difficult to scale up with the increase of actuators. Enabled by imitation learning and DNN feedback, decentralized configuration can also be deployed for performance guaranteed control, resolving the inherent scalability challenge. The frequency responses shown in Fig. \ref{fig:dnn_nonlinear_response} (b) also indicate that the DNN controllers provide near-optimal control signals as both frequency responses remain above and close to the prescribed safety limit.

\section{Conclusion}
This paper presents a stability-constrained imitation learning framework to design DNN-based frequency controller for WTGs to support frequency in low-inertia microgrids. In this framework, the dominant frequency dynamics are captured by a reduced-order ASFR model and then used to generate optimal control signals with frequency nadir constraint for imitation learning. To ensure closed-loop stability, a set of stability-constrained learning formulations is derived using absolute stability and LMIs and solved using the ADMM algorithm.

The proposed method is verified using full-scale nonlinear EMT simulations on the modified IEEE 33-node microgrid. The results demonstrate that the proposed DNN-based controller can let WTGs provide sufficient supportive active power to limit the frequency from dropping below the prescribed threshold under various disturbances while maintaining closed-loop stability. It is shown that a linear feedback control under imitation learning framework can only guarantee frequency safety under a fixed disturbance. Hence, the necessity of using DNN feedback has been demonstrated. 

Furthermore, the DNN controllers provide near-optimal control signals in all three cases, as the frequency responses are not excessively enhanced
but instead remain above and close to the prescribed safety limit.

Finally, this design admits a natural decentralized intelligence. 
As demonstrated, each DNN control can guarantee the frequency performance of its targeted DSG without the need to communicate with other controllers. This feature has a profound meaning as it resolves the long-standing scalability challenge in performance guaranteed control frameworks.

\printbibliography

\end{document}